\documentstyle[12pt]{article}
\def\journal{\topmargin .3in    \oddsidemargin .5in
        \headheight 0pt \headsep 0pt
        \textwidth 5.625in 
        \textheight 8.25in 
        \marginparwidth 1.5in
        \parindent 2em
        \parskip .5ex plus .1ex         \jot = 1.5ex}
\journal

\catcode`\@=11

\catcode`\@=11
\def\section{\@startsection {section}{1}{0pt}{-3.5ex plus -1ex minus
 -.2ex}{2.3ex plus .2ex}{\raggedright\large\bf}}
\catcode`\@=12
\begin{document}
\begin{titlepage}
\begin{center}
November 4, 1993     \hfill    LBL-34704 \\
                     \hfill    UCB-PTH-93/27 \\

\vskip .5in

{\large \bf Baryon decuplet mass relations in chiral perturbation theory}
\footnote{This work was supported in part by the Director, Office of 
Energy Research, Office of High Energy and Nuclear Physics, Division of 
High Energy Physics of the U.S. Department of Energy under Contract 
DE-AC03-76SF00098 and in part by the National Science Foundation under 
grant PHY-90-21139.}

\vskip .5in

Richard F. Lebed
\footnote{lebed@theorm.lbl.gov}
\\[.5in]

{\em  Theoretical Physics Group\\
      Lawrence Berkeley Laboratory\\
      University of California\\
      Berkeley, California 94720}
\end{center}

\vskip .5in

\begin{abstract}
    Baryon decuplet masses within an $SU(3)_L \times SU(3)_R$
chiral Lagrangian formalism are found to satisfy four relations at second order
in flavor breaking.  As a result, the one-loop corrections from Lagrangian
terms up to first order in flavor breaking are observed to give finite and
calculable corrections to these relations.  The formal expressions for these
corrections are presented, followed by numerical evaluations.  We find
consistency between the experimental values of breaking of the relations and the
loop corrections of these relations as predicted by chiral perturbation theory.
\end{abstract}
\end{titlepage}
\renewcommand{\thepage}{\roman{page}}
\setcounter{page}{2}
\mbox{ }

\vskip 1in

\begin{center}
{\bf Disclaimer}
\end{center}

\vskip .2in

\begin{scriptsize}
\begin{quotation}
This document was prepared as an account of work sponsored by the United
States Government.  Neither the United States Government nor any agency
thereof, nor The Regents of the University of California, nor any of their
employees, makes any warranty, express or implied, or assumes any legal
liability or responsibility for the accuracy, completeness, or usefulness
of any information, apparatus, product, or process disclosed, or represents
that its use would not infringe privately owned rights.  Reference herein
to any specific commercial products process, or service by its trade name,
trademark, manufacturer, or otherwise, does not necessarily constitute or
imply its endorsement, recommendation, or favoring by the United States
Government or any agency thereof, or The Regents of the University of
California.  The views and opinions of authors expressed herein do not
necessarily state or reflect those of the United States Government or any
agency thereof of The Regents of the University of California and shall
not be used for advertising or product endorsement purposes.
\end{quotation}
\end{scriptsize}

\vskip 2in

\begin{center}
\begin{small}
{\it Lawrence Berkeley Laboratory is an equal opportunity employer.}
\end{small}
\end{center}

\newpage
\renewcommand{\thepage}{\arabic{page}}
\setcounter{page}{1}
\section{Introduction}
In this paper we extract information about the baryon decuplet mass spectrum
from an effective Lagrangian with $SU(3)_L \times SU(3)_R$ chiral symmetry.
Little
experimental refinement of the decuplet masses has been performed in the past
fifteen years, and decuplet mass differences, particularly isospin splittings,
have large relative errors.  The mass of one decuplet baryon, the $\Delta^-$,
has not even been measured directly, because it cannot be produced as a
low-energy proton-pion resonance.  Therefore, relations among the decuplet
baryons as derived from chiral symmetry and its known breakings (through quark
masses and charges), for the time being, have some predictive power:  one may
simply use chiral relations to rewrite the highly uncertain baryon masses in
terms of their better-measured counterparts.

    Once the decuplet masses have been measured with greater precision, we will
be able to judge the accuracy of the relations.  The amount by which those
combinations of decuplet masses which are predicted to satisfy the relations
fail to do so is a measure of the reliability of the chiral theory we are using
and the size of the contributions of those higher-order chiral terms which do
not satisfy the relations.  We can also compute loop corrections in the chiral
theory to an order consistent with the tree-level calculations and decide
whether the breaking of the relations at tree level can be explained by
lowest-order loop effects.  Thus the careful examination of these relations
provides an important consistency check on the trustworthiness of chiral
perturbation theory.  In this paper we derive four such relations, study their
group-theoretical nature, and find that the one-loop contributions are
consistent with the sizes of the relation breakings, with remaining
discrepancies easily explained by the presence of contributions at third order
in quark masses and charges.
    
   This paper is organized as follows:  In Section 2, we briefly review the
motivation and construction of heavy baryon chiral perturbation theory, which is
the method we utilize to perform computations.  In Section 3, the decuplet
Lagrangian is constructed as an expansion to second order in flavor breaking.
We examine the free parameters of the theory and point out group-theoretical
redundancies, which allow us to count the number of expected mass relations in
Section 4.  The four relations which are found hold to second order in
flavor breaking are exhibited in Section 5, along with group-theoretical reasons
for their generality.  We also present the general formula for decuplet masses
in terms of group-theoretical chiral coefficients.  In Section 6, we
demonstrate that certain finite one-loop corrections to the relations are
calculable, and show why some loop contributions nontrivially vanish.  In
Section 7, we present current numerical results for both the relation breakings
and the full set of chiral coefficients, using various choices of measured
masses, and find that we can explain the amount by which decuplet mass
relations are broken by their lowest-order loop corrections, thus lending
credence to heavy baryon chiral perturbation theory.  We summarize our
conclusions in Section 8.

\section{Heavy baryon theory and the effective Lagrangian}
    An effective field theory which takes into account the approximate $SU(3)
_{L} \times SU(3)_{R}$ chiral symmetry of the three lightest quark flavors and
its spontaneous symmetry breakdown to a more closely satisfied $SU(3)_{V}$
symmetry, namely strong isospin, has long been used to describe successfully
the low-momentum physics of the Goldstone bosons of this symmetry breaking,
the octet of light pseudoscalar mesons\cite{Wein}.  The keys to this
approach are twofold:  first, that the effective Lagrangian, written using
the meson octet fields as the fundamental degrees of freedom, and including the
infinite set of all terms satisfying the same symmetries as the original
Lagrangian (written in terms of the fundamental quark fields), is {\it
formally\/} equivalent to the original Lagrangian; and second, that the terms
in the effective Lagrangian may be organized perturbatively in successively
higher orders of the chiral symmetry-breaking parameters and the derivative
operator (divided by the dimensionally appropriate powers of the chiral symmetry
breaking scale $\Lambda_{\chi} \sim$ 1 GeV), which give calculated contributions
to physical processes that decrease with the order of the terms.  For this
second requirement to be satisfied, it is necessary both that the parameters
which break chiral symmetry explicitly, namely current quark masses and their
electric charges, give contributions sufficiently small (which is guaranteed by
$m_{u,d,s} \ll \Lambda_{\chi}$ and $\alpha_{EM} \ll 1$), and
that the meson momenta are small compared to $\Lambda_{\chi}$.  The latter
condition in particular requires a small meson mass.

    Because the masses of the ground-state baryons are already about 1 GeV, the
assumptions of chiral perturbativity are violated by a chiral Lagrangian naively
including baryons fields as degrees of freedom.  However, an indication of how
one may avoid this problem is suggested by heavy quark effective field
theory\cite{HQT}.  We will use the formalism employed by Georgi\cite{Georgi}.
One considers a chiral baryon multiplet to be a collection of heavy, nearly
on-shell particles degenerate in mass $m_B$ and unit-norm four-velocity
$v^{\mu}$, and having momentum
\begin{equation}
    p^{\mu} = m_B v^{\mu} + k^{\mu} ,
\end{equation}
where $k^{\mu}$ is the residual, off-shell momentum of the baryon.  The
statement that the baryons are nearly on-shell is expressed by the constraint
$k \cdot v \ll m_B$; thus we have a momentum small enough that we
may construct a perturbative chiral theory.  This is implemented by the
transformation on the baryon fields in any multiplet $B$ via
\begin{equation} \label{rot}
    B_v (x) \equiv e^{i m_B \not \, v v_{\mu} x^{\mu}} B(x).
\end{equation}
The positive- and negative-energy solutions are separated by means of the
projection operators
\begin{equation}
    P_{\pm} \equiv \frac{1}{2} (1 \, \pm \not \! v) ,
\end{equation}
and we work only with the positive solutions.  Whereas the original free fields
$B$ satisfy the usual Dirac equation $(i \hspace{-0.4em} \not \hspace{-0.4em}
\partial - m_B)B = 0$, the
new free fields $B_v$ satisfy a massless Dirac equation
\mbox{$i \!  \not  \! \partial B_v = 0$}.  This means that a derivative acting
upon $B_v$ will pull down a factor of $k$ rather than $p$, producing the
perturbative expansion we require.  Henceforth we will work only in
perturbations about these
effectively massless fields, and drop the subscript $v$.  This method, called
heavy baryon effective field theory (HBEFT), has been developed as a useful
calculational tool by Jenkins and Manohar\cite{Jenk1,Jenk2}, although the
general method could be applied to effective field theories with other heavy
degrees of freedom.  However, it must be stressed that we have lost some
information in HBEFT, namely the baryon multiplet mass.  The parameter $m_B$
is nowhere present, so the Lagrangian will be sensitive only to baryon mass
{\it differences}.

    That baryon multiplets are assumed to be degenerate in lowest-order mass and
four-velocity makes it convenient to write a Lagrangian expression ${\cal L}_v$
in the baryon fields for each velocity $v$.  As with the fields, we will
henceforth suppress the index $v$ on ${\cal L}_v$.  The use of this description
has the effect of greatly simplifying the
Dirac algebra; in particular, particles and antiparticles no longer mix,
reducing four-spinors to two-spinors.  It is easily seen\cite{Jenk1} that any
gamma matrix structure in baryon bilinears can be replaced with the $c$-number
velocity $v^{\mu}$ and a generalized spin operator $S^{\mu}$ defined by the
properties
\begin{equation}
\left[ S^{\mu} , S^{\nu} \right]_{+} = \frac{1}{2} \left( v^{\mu} v^{\nu} -
g^{\mu \nu} \right) , \hspace{2em} \left[ S^{\mu} , S^{\nu} \right]_{-} = i
\epsilon^{\mu \nu \rho \sigma} v_{\rho} S_{\sigma} ,
\end{equation}
where $\epsilon_{0123} = + 1$.  A specific representation of the operator
satisfying these relations is
\begin{equation}
S^{\mu} = \frac{1}{2} \left( \not \! v \gamma^{\mu} - v^{\mu} \right)
\gamma_{5} .
\end{equation}
In the rest frame ($v^{\mu} = (1, \bf 0)$),
the operator $S^{\mu}$ is found to reduce to the usual Pauli matrices $(0,
\mbox{\boldmath $\sigma$}/2)$ with the usual commutation relations.

The prescriptions listed above are useful for any baryon multiplet included in
our theory; in the current model we must consider both baryon decuplet and
octet degrees of freedom.  Whereas the octet baryon fields are taken to be
ordinary Dirac fields, the decuplet is taken to be represented by
Rarita-Schwinger fields $\, T^\mu_{ijk}$, symmetric under permutations of the
$SU(3)$ flavor indices $i,j,k$, with the spin-1/2 portion projected out via the
usual constraint $\gamma_{\mu} T^{\mu} = 0$.  In HBEFT, this translates to the
two conditions
\begin{eqnarray}
    v_{\mu} T^{\mu} = 0, & & S_{\mu} T^{\mu} = 0 .
\end{eqnarray}
The consequences of these conditions for the Feynman rules of this theory are
summarized in Reference~\cite{Jenk2}.

One point, however, that should be made at this time is that the octet and
decuplet in this theory are taken to have different multiplet common masses,
$m_B = m_8, m_{10}$.  Both fields, $B^i{}_j$ and $T^{\mu}_{ijk}$, transform
under the rule given by Eq.~\ref{rot}; since the values of $m_B$ in the phases
are different, the intermultiplet spacing $\Delta m \equiv m_{10} - m_8$ is a
parameter which appears in the Lagrangian and must be placed into the theory by
hand.

\section{Constructing the Lagrangian}
\subsection{Field transformation properties}
The hadron field multiplets may be compactly represented by matrices in flavor
space.  The baryon and meson octets have the familiar forms
\begin{equation}
B = \left( \begin{array}{ccc} \label{m1}
\frac{1}{\sqrt{2}} \Sigma^{0} + \frac{1}{\sqrt{6}} \Lambda & \Sigma^{+} & p \\
\Sigma^{-} & -\frac{1}{\sqrt{2}} \Sigma^{0} + \frac{1}{\sqrt{6}} \Lambda & n \\
\Xi^{-} & \Xi^{0} & -\frac{2}{\sqrt{6}} \Lambda
\end{array} \right) ,
\end{equation} 
and
\begin{equation}
\Pi = \frac{1}{\sqrt{2}} \left( \begin{array}{ccc}
\frac{1}{\sqrt{2}} \pi^{0} + \frac{1}{\sqrt{6}} \eta & \pi^{+} & K^{+} \\
\pi^{-} & -\frac{1}{\sqrt{2}} \pi^{0} + \frac{1}{\sqrt{6}} \eta & K^{0} \\
K^{-} & \overline{K}^{0} & -\frac{2}{\sqrt{6}} \eta
\end{array} \right) .
\end{equation} 
The baryon decuplet in this notation, a $3 \times 3 \times 3$ array, may be
represented as (suppressing Lorentz indices) a collection of three matrices:
\begin{eqnarray} \label{m3}
\lefteqn{T_{ijk} =} \nonumber \\ & &
    \hspace{-2em}
    \left( 
    \hspace{-0.5em}
    \begin{array}{ccc}
\Delta^{++} & \frac{1}{\sqrt{3}} \Delta^{+} & \frac{1}{\sqrt{3}} \Sigma^{*+} \\
\frac{1}{\sqrt{3}} \Delta^{+} & \frac{1}{\sqrt{3}} \Delta^{0} &
\frac{1}{\sqrt{6}} \Sigma^{*0} \\
\frac{1}{\sqrt{3}} \Sigma^{*+} & \frac{1}{\sqrt{6}} \Sigma^{*0} &
\frac{1}{\sqrt{3}} \Xi^{*0}
    \end{array}
    \hspace{-0.5em}
    \right)
    \hspace{-0.5em}
    \left(
    \hspace{-0.5em}
    \begin{array}{ccc}
\frac{1}{\sqrt{3}} \Delta^{+} & \frac{1}{\sqrt{3}} \Delta^{0} &
\frac{1}{\sqrt{6}} \Sigma^{*0} \\
\frac{1}{\sqrt{3}} \Delta^{0} & \Delta^{-} & \frac{1}{\sqrt{3}} \Sigma^{*-} \\
\frac{1}{\sqrt{6}} \Sigma^{*0} & \frac{1}{\sqrt{3}} \Sigma^{*-} &
\frac{1}{\sqrt{3}} \Xi^{*-}
    \end{array}
    \hspace{-0.5em}
    \right)
    \hspace{-0.5em}
    \left(
    \hspace{-0.5em}
    \begin{array}{ccc}
\frac{1}{\sqrt{3}} \Sigma^{*+} & \frac{1}{\sqrt{6}} \Sigma^{*0} &
\frac{1}{\sqrt{3}} \Xi^{*0} \\
\frac{1}{\sqrt{6}} \Sigma^{*0} & \frac{1}{\sqrt{3}} \Sigma^{*-} &
\frac{1}{\sqrt{3}} \Xi^{*-} \\
\frac{1}{\sqrt{3}} \Xi^{*0} & \frac{1}{\sqrt{3}} \Xi^{*-} & \Omega^{-}
    \end{array}
    \hspace{-0.5em}
    \right) . \nonumber \\ & &
\end{eqnarray}
One may assign any particular permutation of indices {\it i,j,k\/} to denote 
row, column, and sub-matrix in this representation, because the decuplet is
completely symmetric under rearrangement of flavor indices. 

    We require in particular that the baryon chiral Lagrangian contains the
usual nonlinear sigma model.  To this end, we define the fields
\begin{equation}
\xi \equiv e^{i \Pi / f} , \hspace{2em} \Sigma \equiv \xi^2 = e^{2i \Pi / f} ,
\end{equation}
where the choice of pion decay constant normalization is $f  \approx 93$ MeV.
Then, with $L$ and $R$ specifying the left- and right-handed chiral
transformations respectively, the field $\xi$ is chosen to transform under
$SU(3)_L \times SU(3)_R$ according to the rule
\begin{equation}
\xi \mapsto L \xi U^{\dagger} =  U \xi R^{\dagger} ,
\end{equation}
a mapping which implicitly defines the transformation $U$, and which implies the
usual transformation
\begin{equation}
\Sigma \mapsto L \Sigma R^{\dagger} .
\end{equation}
Under our transformation choice of $\xi$, we may define Hermitean vector and
axial vector currents:
\begin{equation}
V^{\mu} \equiv \frac{i}{2} \left( \xi \partial^{\mu} \xi^{\dagger} +
\xi^{\dagger} \partial^{\mu} \xi \right) , \hspace{2em} A^{\mu} \equiv
\frac{i}{2} \left( \xi \partial^{\mu} \xi^{\dagger} - \xi^{\dagger}
\partial^{\mu} \xi \right) .
\end{equation}
So that these currents may couple to the baryon fields in a chirally-invariant
way, we need the baryonic transformation properties
\begin{equation} \label{xform}
B \mapsto U B U^{\dagger} , \hspace{2em}
T^{\mu}_{ijk} \mapsto U^l{}_i U^m{}_j U^n{}_k T^{\mu}_{lmn},
\end{equation}
and the chirally-covariant derivatives
\begin{eqnarray}
{\cal D}^{\nu} B & \equiv & \partial^{\nu} B - i \left[ V^{\nu} , B \right]_{-},
\nonumber \\ \left( {\cal D}^{\nu} T^{\mu}\right)_{ijk} & \equiv &
\partial^{\nu}
T^{\mu}_{ijk} - i (V^{\nu})_i{}^l T^{\mu}_{ljk} - i (V^{\nu})_j{}^l
T^{\mu}_{ilk} - i  (V^{\nu})_k{}^l T^{\mu}_{ijl} ,
\end{eqnarray}
which have the same transformation properties (Eq.~\ref{xform}) as the baryon
fields upon which they act. 

The explicitly chiral symmetry-breaking operators, namely
\begin{equation}
    M_q \equiv \left( \begin{array}{ccc}
m_u & & \\ & m_d & \\ & & m_s
    \end{array} \right) , \hspace{2em}
    Q_q \equiv \left( \begin{array}{ccc}
+\frac{2}{3} & & \\ & -\frac{1}{3} & \\ & & -\frac{1}{3}
    \end{array} \right) ,
\end{equation}
are included by the usual spurion procedure, treating them as fields which
transform under $SU(3)_L \times SU(3)_R$ as the conjugate representations of the
amount by which they break the chiral symmetry of the QCD Lagrangian.  The
relevant pieces of the QCD Lagrangian in a chiral basis are
\begin{equation}
\delta {\cal L} = -\left( \overline{\psi}_L M_q \psi_R + \mbox{h.c.} \right) - e
A^{\mu} \left( \overline{\psi}_L Q_L \gamma_{\mu} \psi_L + \overline{\psi}_R Q_R
\gamma_{\mu} \psi_R \right) ,
\end{equation}
where
\begin{eqnarray}
\psi \equiv \left( \begin{array}{c} u \\ d \\ s \end{array} \right) , &
Q_L \equiv Q_R \equiv Q_q .
\end{eqnarray}
From this we obtain the usual rules
\begin{eqnarray}
M_q & \mapsto & L M_q R^{\dagger} , \\
Q_L & \mapsto & L Q_L L^{\dagger} , \\
Q_R & \mapsto & R Q_R R^{\dagger} .
\end{eqnarray}
One can quickly see in the theory with the field $\xi$ that the matrices $M_q$
and $Q_q$ will always occur in the combinations
\begin{eqnarray}
M_{+} \equiv \frac{1}{2} \left( \xi^{\dagger} M_q \xi^{\dagger} + \xi M_q
\xi \right), & & M_{-} \equiv \frac{i}{2} \left( \xi^{\dagger} M_q
\xi^{\dagger} - \xi M_q \xi \right), \\ Q_{+} \equiv \frac{1}{2} \left(
\xi^{\dagger} Q_L \xi + \xi Q_R \xi^{\dagger} \right) , & & \, Q_{-} \equiv
\frac{i}{2} \left( \xi^{\dagger} Q_L \xi - \xi Q_R \xi^{\dagger} \right) ,
\end{eqnarray}
which are designed to be Hermitean, have definite parity properties as indicated
by the subscript, and transform appropriately under $SU(3)_L \times SU(3)_R$:
\begin{equation}
X \mapsto U X U^{\dagger}, \mbox{   for $X = A^{\mu}, M_{\pm}, Q_{\pm}$.}
\end{equation}

\subsection{Lagrangian terms}
Now it is a simple matter to construct the most general Lagrangian.  This
model is constructed to include all terms to two orders in the perturbation
operators we have discussed, namely all terms with a total of two of the
following operators: $\partial^{\mu}$, $M_q$, and $Q_q$.  Because the derivative
operators ultimately generate meson masses and hence quark masses, the series
may be thought of as one in quark masses and charges alone.  The physics of the
expansion becomes more lucid when we distinguish our results by the number of
powers of $m_s$, $m_{u,d}$, and $\alpha_{EM}$, namely, organizing them
according to their $SU(3)$ and isospin-$SU(2)$ properties.  Since we are
representing strong and electromagnetic interactions only, we include only
those terms that respect the same symmetries.  Charge conjugation symmetry
eliminates all terms with an odd total number of $Q_{\pm}$, whereas parity
conservation requires that $M_{-}$ and $Q_{-}$ only occur in an even total
number (before we include derivative terms), that is, not before second order.
Furthermore, since $M_{-}$ and $Q_{-}$, when expanded in powers of the meson
fields, give no contribution before $o(\Pi^{2})$, these terms would not appear
in mass relations until we included loops of {\it second\/}-order
terms.  Thus the parity-odd combinations do not occur at this order, and we may
therefore suppress the subscript ($+$).
    
Next, all terms with derivatives produce factors of meson momentum and thus
contribute to masses only through diagrams with meson loops.  Because
higher-order loops require more meson fields and hence produce more powers of
the
quark masses, we compute only those loop diagrams with one meson
loop.  Derivatives appear in the Lagrangian through the covariant derivative
${\cal D}^{\mu} (= \partial^{\mu} + o(\Pi^2))$ and through the axial vector
current $A^{\mu} (= \partial^{\mu} \Pi /f + o(\Pi^3))$.  Meson fields also
occur in $M = M_q + o(\Pi)^2$ and $Q = Q_q + o(\Pi^2)$.  We quickly learn that
all one-meson loop diagrams at one derivative will either be of the ``keyhole''
variety (Figure 1a) or have two separated insertions of $A^{\mu}$ (Figure 1b).

There is a bewildering proliferation of terms at second order once we include
derivative terms, for example,
\begin{eqnarray}
\overline{T} (v \cdot A) (S \cdot{\cal D}) T, & & \overline{T} (v \cdot A) M_{-}
T,
\end{eqnarray} 
and many others.  Two constraints, however, simplify the situation:  the first
is that we are only computing diagrams with one meson loop; and the second is
that terms with one covariant derivative $\cal D$ and some other operator $X$
can be transformed away by a suitable redefinition of the baryon field and the
addition of new $o(X^2)$ terms.  The physical reasoning behind this
transformation is that the only one-derivative terms we allow in the fermion
Lagrangian are the kinetic terms; if we find other one-derivative terms, it
means that we will have non-canonical equations of motion for the baryons and
unfamiliar forms for the Feynman propagators.  Under these rules, the only
remaining terms at second order and including meson fields are of the forms
\begin{eqnarray} \label{2nd}
\overline{T} A A T, & & \overline{T} {\cal D} {\cal D} T,
\end{eqnarray}
with appropriate factors of $v$ and $S$ thrown in to give the terms the correct
Lorentz structure.  Clearly, the only one-loop graphs possible from these terms
are keyhole diagrams; this will prove to have special significance when we
consider loop corrections to certain mass combinations.

The light-flavor symmetry $SU(3)_V$, which includes strong isospin, is broken
in this model only by the inequality of quark masses and charges; all other
operators and
coefficients are assumed to obey chiral symmetry.  Under these restrictions,
the most general decuplet Lagrangian we need to consider in this model is
\begin{eqnarray} \label{lag}
{\cal L} & = &
\mbox{} -i \overline{T}^{\mu}_{ijk} v_{\nu} \left( {\cal D}^{\nu} T_{\mu}
\right)^{ijk} +
\Delta m \overline{T}^{\mu}_{ijk} T_{\mu}^{ijk} + 2 {\cal H}
\overline{T}^{\mu}_{ijk} S_{\nu} \left( A^{\nu} \right) ^k{}_l T_{\mu}^{ijl}
\nonumber \\ & &
\mbox{} - 2 \tilde{\sigma} M^l{}_l \overline{T}^{\mu}_{ijk} T_{\mu}^{ijk} + 2c
\overline{T}^{\mu}_{ijk} M^k{}_l T_{\mu}^{ijl} \nonumber \\ & &
\mbox{} + \frac{\alpha}{4 \pi} \Lambda{\chi} \left\{ d_1 Q^l{}_l Q^m{}_m
\overline{T}^{\mu}_{ijk} T_{\mu}^{ijk} + d_2 Q^l{}_m Q^m{}_l
\overline{T}^{\mu}_{ijk} T_{\mu}^{ijk} \right. \nonumber \\ & &
\left. \mbox{} \hspace {4em} + f_1 \overline{T}^{\mu}_{ijk} Q^k{}_l Q^l_m
T_{\mu}^{ijm} + f_2 \overline{T}^{\mu}_{ijk} Q^j{}_l Q^k{}_m T_{\mu}^{ilm}
\right\} \nonumber \\ & &
\mbox{} + \frac{1}{\Lambda{\chi}} \left\{ g_1 M^l{}_l M^m{}_m
\overline{T}^{\mu}_{ijk} T_{\mu}^{ijk} + g_2 M^l{}_m M^m{}_l
\overline{T}^{\mu}_{ijk} T_{\mu}^{ijk} \right. \nonumber \\ & &
\left. \hspace{3em} \mbox{} + k_1 \overline{T}^{\mu}_{ijk} M^k{}_l M^l_m
T_{\mu}^{ijm} + k_2 \overline{T}^{\mu}_{ijk} M^j{}_l M^k{}_m T_{\mu}^{ilm}
\right\} \nonumber \\ & &
\mbox{} + \mbox{terms of the form Eq.~\ref{2nd}} .
\end{eqnarray}
The sign of the kinetic term follows from the fact that the Rarita-Schwinger
spinor solutions are spacelike.  The coefficients $c$, $\tilde{\sigma}$,
$d_{1,2}$, $f_{1,2}$, $g_{1,2}$, and $k_{1,2}$ are normalized to be unitless and
expected to be of order one (as demanded by naturalness).  The factor
$\frac{\alpha}{4 \pi}$ multiplying the electromagnetic terms follows from the
fact that these terms, if computed from the quark Lagrangian, would arise from
photon loop diagrams. 

\section{Parameter counting}
A naive counting of the parameters in the Lagrangian, Eq.~\ref{lag}, gives us
one decuplet-octet mass breaking $\Delta m$, two first-order coefficients in $M$
($\tilde{\sigma}$ and $c$), four at second order in each of $o(M^2)$ ($g_{1,2}$
and $k_{1,2}$) and $o(Q^2)$ ($d_{1,2}$ and $f_{1,2}$), and the three light quark
masses $m_{u,d,s}$.  The coefficients of the derivative terms can be measured,
at least in principle, in decuplet decay processes.  This gives a total of
fourteen fit parameters, and the situation appears to be hopeless.  However,
not all of the coefficients and quark masses in the Lagrangian are independent
parameters.  Many can be eliminated by a careful examination of the
group-theoretical properties of the various terms.

We first consider those terms which appear as singlets in the Lagrangian.  In
a more usual Rarita-Schwinger theory, the mass term is just
\begin{equation}
\delta {\cal L} = m_{10} {\overline T}^{\mu}_{ijk} T_{\mu}^{ijk} .
\end{equation}
This term does not occur in HBEFT, for we have eliminated the common mass term
through field redefinition.  If a term which is of the same form appears at
higher order in perturbation theory, that is, an operator singlet multiplying
the fully contracted ${\overline T}T$ fields, it too will contribute only to the
overall multiplet mass, and thus may have been defined away at the outset.  The
terms in this category are those with coefficients $\Delta m$, $d_{1,2}$, and
$g_{1,2}$.  Note, however, that $\Delta m$ can be neglected only as long as
there is only one multiplet mass in the theory.

The next observation applies equally to any chiral perturbation theory which
takes any set of parameters as undetermined inputs; we will consider the
particular case of the current quark masses.  Each term in the chiral
Lagrangian then
contains a certain number of factors of the quark mass matrix  $M_q$.  The
coefficients $c_i$ of the various terms are {\it a priori} unrelated parameters
(unlike in a renormalized field theory).  Given a particular term with $n_i$
powers of $M_q$, one readily sees that the term is invariant under the
transformation
\begin{equation}
M_q \mapsto k M_q , \hspace{2em} c_i \mapsto k^{-n_i} c_i ,
\end{equation}
where $k$ is arbitrary.  Thus, in any chiral theory, one cannot hope to obtain
quark masses, but only ratios of quark masses.  For three light flavors, we have
two ratios; we will find it convenient to use two particular such ratios,
\begin{equation}
q,r \equiv \frac{m_d \pm m_u}{m_s - \frac{1}{2} (m_u + m_d)} .
\end{equation}
Note that both parameters are small, inasmuch as $m_{u,d} \ll m_s$, and that $q$
can appear with isospin-conserving operators, whereas $r$ can only appear with
isospin-breaking operators.

We can follow this argument one step farther in HBEFT.  We simply observe that
the tree-level Lagrangian is insensitive to transformations of the type
$M_q \mapsto M_q +
c {\bf 1}$, where {\bf 1} is the identity matrix and $c$ is arbitrary.  This
follows because each insertion of {\bf 1} in a term of $o(M_q^n)$ is equivalent
to a redefinition of the coefficient of the terms of lower orders in $M_q$ by
simple binomial expansion.  Eventually, we generate singlet terms in this way,
which we may ignore, using the argument described above.  In particular, this
tells us that the Lagrangian is sensitive only to differences of quark masses;
or, combining this with the previous result, it is sensitive only to
{\it ratios of differences} of quark masses.  For three light flavors, only one 
parameter remains, which we choose to be the parameter $r$.

Because the relevant operator is not merely the number matrix $M_q$,
but the full operator $M = M_q + o(\Pi^2)$, the coefficients no longer shift
simply by $c$-numbers.  The above argument then fails to hold, and we must use
both quark mass parameters, $q$ and $r$.  However, this situation is true only
for those terms for which we must include the meson fields in loops.  In our
expansion, this means the first-order terms in $M$.  The
tree-level masses can always be written using only $r$.  Because, as described
in the Appendix, a diagram of the form of Figure 1a may be written in terms of
the internal baryon mass, the only diagrams in this model in which $q$ would
appear are the keyhole diagrams of the operator $M$; however, as we will see in
Section 6, such contributions cancel in the relations we derive.

There is still one more major simplification of the Lagrangian.  Consider any $3
\times 3$ matrix $X$ as an operator in the mass contribution
\begin{equation}
\overline{T}^{\mu}_{ijk} X^k{}_l T_{\mu}^{ijl} .
\end{equation}
Because $T$ is symmetric in its flavor indices, the placement of the contraction
is irrelevant.  Group-theoretically, $X$ has a singlet part, which we may
ignore, and an octet part.  The raising and lowering operators included in $X$,
which have the off-diagonal matrix entries as coefficients, do not contribute to
mass terms, which are diagonal.  This leaves only operators proportional to the
Cartan subalgebra of isospin and hypercharge.  No matter how many operators we
have of the form of $X$, they can always be redefined into only two:  one
proportional to $T^3$ and one proportional to $T^8$.  These operators will obey
equal-spacing rules in isospin and hypercharge.  We have three such operators
in our model: those with coefficients $c$, $f_1$, and $k_1$.  A more careful
analysis of the group theory of decuplet masses will appear in the next Section.

We are again ready to count parameters. There are the two operators proportional
to isospin and hypercharge; one quark mass parameter $r$, and one remaining
$o(M^2)$ and one $o(Q^2)$ term, making five unknowns.  On the other side of
the equation, we begin with ten decuplet masses, but again, HBEFT tells us that
we can discover nothing of the common decuplet mass.  We thus have nine mass
differences, implying that there are four nontrivial mass relations
between the decuplet baryons.

\section{Decuplet mass relations}
Using the redefinition properties of the last Section, we find that there are
exactly four mass relations between the decuplet masses that hold to
second-order in flavor-breaking in the tree-level Lagrangian.  In a
symmetric form, they may be written as
\begin{eqnarray}
0 & \hspace{-0.5em} = & \hspace{-0.5em} \Delta^{++} - 3 \Delta^{+} + 3
\Delta^{0} - \Delta^{-} , \label{first} \\
0 & \hspace{-0.5em} = & \hspace{-0.5em} \left( \Delta^{++} - \Delta^{+} -
\Delta^{0} + \Delta^{-} \right) - 2 \left( \Sigma^{*+} - 2 \Sigma^{*0} +
\Sigma^{*-} \right) , \\
0 & \hspace{-0.5em} = & \hspace{-0.5em} \left( \Delta^{+} - \Delta^{0} \right)
- \left( \Sigma^{*+} - \Sigma^{*-} \right) + \left( \Xi^{*0} - \Xi^{*-} \right),
\\
0 & \hspace{-0.5em} = & \hspace{-0.5em} \frac{1}{4} \left( \Delta^{++} +
\Delta^{+} + \Delta^{0} + \Delta^{-} \right) - \left( \Sigma^{*+} +
\Sigma^{*0} + \Sigma^{*-} \right) + \frac{3}{2} \left( \Xi^{*0} + \Xi^{*-}
\right) - \Omega^{-} . \nonumber \label{last} \\ & &
\end{eqnarray}
Notice that the first three of these are isospin-breaking, and only the fourth
remains in the limit that isospin is a good symmetry.  We label the amounts by
which these relations are experimentally broken as $\Delta_{1,2,3,4}$,
respectively.

These relations are not unknown in the literature.  In fact, the
first three can be trivially derived from quark model calculations over a
quarter of a century old\cite{IKS,GS}.  The original form of the relations
connects the masses of the octet to decuplet baryons, for when these were first
derived, the octet masses were much better known than the decuplet masses.
This does not occur in the form of HBEFT that we use, as the octet and decuplet
are treated as two independent multiplets, unrelated by the physical fact that
they are both three-quark states, and the multiplet mass splitting $\Delta m$
is an independent parameter.

Eq.~\ref{last} is the remnant of the Gell-Mann's famous equal-spacing rule,
\begin{equation}
\Delta - \Sigma^{*} = \Sigma^{*} - \Xi^{*} = \Xi^{*} - \Omega .
\end{equation}
These are readily obtained from the Lagrangian, Eq. \ref{lag}, if we set the
isospin-violating parameter $r$ to zero and neglect second-order terms.  It was
pointed out by Okubo\cite{Okubo} as early as 1963 that, to second order in
flavor breaking, the only surviving such relation is
\begin{equation}
\Delta - 3 \Sigma^* + 3 \Xi^* - \Omega = 0,
\end{equation}
which is, neglecting isospin states, the same as Eq.~\ref{last}.  It was also
derived in this form by Jenkins\cite{Jenk3}, using HBEFT.

Equivalent forms of all four relations have been derived in a very general quark
model, the general parametrization method of Morpurgo\cite{Morp}.
Again, the relations were written in terms of equations connecting octet to
decuplet masses.

Let us examine the reasons that Eqs.~\ref{first}--\ref{last} hold.
Eq.~\ref{first} is simply a consequence of the Wigner--Eckart theorem applied
to an isospin-3/2 multiplet for which all Lagrangian mass terms transform as
$\Delta I = 0, 1, 2$, which is the case for our model Lagrangian, Eq.~\ref{lag}.

One can find by direct substitution that, if we replace the matrices $M_{q}$ and
$Q_{q}$ with an {\it arbitrary} $3 \times 3$ matrix $X$, the relations are still
true.  This is as much as saying that the relations hold for each Lagrangian
term first- or second-order in any flavor singlet or octet operator.  Consider
the $SU(3)$ structure of the operator $\cal O$ in an arbitrary decuplet mass
term:
\begin{equation}
\overline{T}^{\mu}_{ijk} {\cal O}^{ijk}{}_{lmn} T_{\mu}^{lmn} .
\end{equation}
Since 
\begin{equation}
{\bf 10} \otimes {\overline {\bf 10}} = {\bf 64} \oplus {\bf 27} \oplus {\bf 8}
\oplus {\bf 1} ,
\end{equation}
and all of these representations are real, the operator ${\cal O}$ must live in
some combination of the representations in the direct sum above.  On the other
hand,
\begin{equation}
{\bf 8} \otimes {\bf 8} = {\bf 27} \oplus {\bf 10} \oplus
{\overline {\bf 10}} \oplus {\bf 8} \oplus {\bf 8} \oplus {\bf 1},
\end{equation}
and so the product of two $3 \times 3$ matrices can produce decuplet mass terms
that transform as ${\bf 1}$, ${\bf 8}$, and ${\bf 27}$.  That we obtain the
result for arbitrary ${\cal O}^{ijk}{}_{lmn} = \delta^{i}{}_{l}
\delta^{j}{}_{n} X^{k}{}_{n}$ means that the relations hold for arbitrary
${\bf 1}$'s and ${\bf 8}$'s; but the ${\bf 27}$ was constructed only from two
identical matrices $X$, and thus is not an arbitrary ${\bf 27}$.

However, we may construct the most general operator containing a ${\bf 27}$,
${\cal O}^{ijk}{}_{lmn} = \delta^{i}{}_{l} X^{jk}{}_{mn}$, and by direct
computation show that in a decuplet mass term it nevertheless respects
Eqs.~\ref{first}--\ref{last}.  This will be extremely useful when we consider
loop corrections to the relations, because certain loop corrections are of this
form and thus do not contribute to mass relation breaking.  We conclude that
only operators transforming under a {\bf 64} will break the relations. 

We may also catalogue the {\it complete\/} $SU(3)$ group-theoretical structure
of the decuplet mass terms independent of their physical origins in the chiral
Lagrangian.  The physical decuplet masses alone determine the numerical size of
these coefficients, and thus limit the size of contributions from terms in the
Lagrangian not appearing in this model.  Denoting the chiral coefficient of
the term with operators transforming under an $R$-dimensional representation of
$SU(3)$ and a $(2I+1)$-dimensional representation of isospin by $c^R_I$, a
straightforward exercise in $SU(3)$ Clebsch-Gordan coefficients gives the
general mass formula
\begin{eqnarray} \label{mass}
M & = & c^1_0 + c^8_0 Y + c^8_1 I_3 + c^{27}_0
\left( 5Y^2 + 3Y - 5 \right) \nonumber \\ & & \mbox{} +
c^{27}_1 I_3 \left( 5Y - 3 \right) +
\frac{1}{4} c^{27}_2 \left( 12I_3^2 - Y^2 -6Y -8 \right) \nonumber \\ & &
\mbox{} + \frac{1}{6} c^{64}_0 \left( 35Y^3 + 45Y^2 - 50Y -24 \right) + c^{64}_1
I_3 \left( 21Y^2 -9Y - 10 \right) \nonumber \\ & & \mbox{} +
\frac{1}{12} c^{64}_2 \left[ 12I_3^2 \left( 7Y - 6 \right) - 7Y^3 - 36Y^2 - 20Y
+ 48 \right] \nonumber \\ & & \mbox{} + \frac{1}{6} c^{64}_3 I_3 \left( 20I_3^2
- 3Y^2 - 18Y - 20 \right) ,
\end{eqnarray}
where hypercharge $Y$ is normalized by $Q = I_3 + \frac{1}{2} Y$.  (Note that
the coefficients are independent of the isospin Casimir $I(I+1)$ for the
decuplet because all states are singly degenerate, so that $I_3$ and $Y$
uniquely determine the state.)  In this notation we find the relation-breaking
parameters to be given by
\begin{eqnarray}
\Delta_1 & = & 20 c^{64}_3 , \\
\Delta_2 & = & 28 c^{64}_2 , \\
\Delta_3 & = & 42 c^{64}_1 - 6 c^{64}_3, \\
\Delta_4 & = & 35 c^{64}_0 .
\end{eqnarray}

An amusing implication of the insensitivity of Eqs.~\ref{first}--\ref{last} to
the form of the chiral breaking, combined with the symmetry of the decuplet
field under permutation of flavor indices, is that the four relations are
invariant under permutations of the three isospin axes $T_3, U_3, V_3$ in the
weight diagram of the decuplet in flavor space; for example, Eq.~\ref{first}
becomes
\begin{equation}
0 = \Delta^{-} - 3 \Sigma^{*-} + 3 \Xi^{*-} - \Omega^{-} ,
\end{equation}
under $T_3 \mapsto U_3$, $U_3 \mapsto -V_3$, $V_3 \mapsto -T_3$, or
\begin{equation}
0 = \Omega^{-} - 3 \Xi^{*0} + 3 \Sigma^{*0} - \Delta^{++} ,
\end{equation}
under $T_3 \mapsto -V_3$, $U_3 \mapsto T_3$, $V_3 \mapsto -U_3$.  That is, the
relations are unaffected if we permute, for example, the up and strange quarks
in all decuplet wavefunctions.

\section{Loop corrections}
As mentioned in the previous Section, the tree-level decuplet mass relations
holding at second order in flavor breaking are not exact, being broken only by
those decuplet mass operators having some portion which transforms under a
$\bf{64}$ representation of $SU(3)$.  Relations which hold to second-order in
quark masses are particularly interesting, because any loop corrections which
have the form $o(m_q^0, m_q^1,\mbox{ or } m_q^2)$ are absorbed through the usual
renormalization procedure by the appropriate counterterms, and therefore must
also satisfy the relations.  As a consequence, the loop corrections below
$o(m_q^3)$ are calculable and nonanalytic in quark masses, and independent of
the arbitrary renormalization point $\mu$.  In one-loop
corrections of the terms with less than two powers of $M$, all corrections are
of this form.  Thus, in this model, one-loop diagrams give calculable
corrections to the relations Eqs.~\ref{first}--\ref{last} if and only if they
contain a piece transforming under a ${\bf 64}$.  An
analogous situation occurs for the octet baryons, as the HBEFT Lagrangian is
found to satisfy one relation at second order\cite{LL}, the Coleman--Glashow
relation.  Here we wish to consider the various meson loop corrections to
decuplet masses and classify them by their transformation properties.    

The one-loop diagrams fall into two basic categories:  those with one quartic
vertex (Figure 1a), which we have called keyhole diagrams, and those with two
trilinear vertices (Figure 1b).  Keyhole diagrams in this model occur in two
forms:  those from the meson-field expansion of the operator $M$ (which
transforms as singlet plus octet, and therefore satisfies the relations), and
those with the structure
\begin{equation} \label{key}
f( \alpha ) \overline{T} \, \Pi^{\alpha} \Pi^{\alpha} T ,
\end{equation}
where flavor indices can appear in all possible contractions, $\alpha$ is the
meson octet index, and $f(\alpha)$ is that function which appears upon
integration over meson momentum.  That the two meson field indices are the same
simply implies that the meson loop must close; $\eta$-$\pi^0$ mixing does not
change this result, because we may rediagonalize the meson $SU(3)$ generators so
that they refer now to mass eigenstates.  All that is important is that both the
unrotated and rotated generators are octet.  Note also that this result
holds for {\it each\/} value of $\alpha$, not just the sum.  Because the largest
representation obtained from the two octet fields is a $\bf{27}$, such loop
expressions satisfy Eqs.~\ref{first}--\ref{last}; thus all keyhole diagrams
satisfy the second-order relations.

Next consider loop diagrams with two trilinear vertices.  In this model, the
intermediate state may be either decuplet or octet, the latter made possible
through the inclusion of the interaction
\begin{equation}
\delta {\cal L} = {\cal C} \left( \epsilon^{ilm} \overline{T}^{\mu}_{ijk}
\left( A^{\mu} \right)^j{}_l B^k{}_m + \mbox{h.c.} \right) .
\end{equation}
For the first time we need to consider the inclusion of the octet baryons in the
decuplet Lagrangian.  That octet and decuplet baryons should be considered
together is a physical statement of the fact that chiral Lagrangians distinguish
neither multiplet as being the more fundamental, because both arise from the
$SU(3)$ product $\bf{3} \otimes \bf{3} \otimes \bf{3}$, and it is only the
choice of the sign of $\Delta m$ that determines which multiplet is heavier.

First, however, consider the case of the diagram with an  intermediate decuplet
baryon.  The general structure of the diagram in this case is
\begin{equation}
{\cal H}^2 f( \alpha ) \sum_{j} \left( \overline{T} \, \Pi^{\alpha} T_{j}
\right) {\cal O}_j \left( \overline{T}_j \Pi^{\alpha} T \right) ,
\end{equation}
where $j$ refers to the ten possible intermediate decuplet states, and $\cal O$
is an operator that may appear on the intermediate line (for example, the $o(M)$
tree-level term).  Now if $\cal O$ is arbitrary, there is little that can be
said using only group theory.  However, if we consider only those diagrams for
which ${\cal O} = \bf{1}$, then we see that the $j$-dependence is trivial, for
then we have a completeness relation over decuplet generators:
\begin{equation}
\sum_{j} T_{j} \overline{T}_j \propto {\bf 1} .
\end{equation}
We find that these diagrams have exactly the same group-theoretic structure as
in Eq.~\ref{key} and therefore also satisfy the relations
Eqs.~\ref{first}--\ref{last}.  Since each loop contributes a factor of
$(16 \pi^2 f^2)^{-1}$, dimensional analysis shows these loops to have quark
mass dependence $o(m_q^{3/2})$.

The diagrams with internal octet baryons have an analogous property:  if the
octet baryons are taken degenerate in mass (but the decuplet masses are
unconstrained), we again have a completeness sum (this time over octet
generators), and conclude that diagrams with ${\cal O} = \bf{1}$ satisfy the
decuplet relations.
However, there is a complication in the presence of the intermultiplet spacing
$\Delta m$: neglecting the mass splittings within the baryon octet seems
unreasonable when many of these splittings are comparable in size to $\Delta m$.
The most pedantic way of including this information would be to solve
simultaneously for the octet chiral coefficients using the experimental octet
mass values, and to include the corresponding operators $\cal O$ in the loop
diagrams.  However, it is more direct to use the octet masses directly; the
expressions and their contributions to the breaking of
Eqs.~\ref{first}--\ref{last} are exhibited in the Appendix.

The remaining decuplet-intermediate diagrams (those with ${\cal O} \neq {\bf
1}$) may also be rewritten in terms of the decuplet baryon masses, rather than
the chiral coefficients; because the loop coefficients are proportional to
powers of quark masses and charges, the difference between using chiral
coefficients and decuplet masses within loops is higher order still
(at least $o(m_q^3 \ln m_q)$) in quark masses and charges.  It is thus not
unreasonable to present the mass expressions in this manner.  These finite
corrections to Eqs.~\ref{first}--\ref{last} are also discussed in the Appendix.

\section{Results and predictions}
We are now ready to test the validity of the relations
Eqs.~\ref{first}--\ref{last} using the expressions listed in the Appendix.
Because, as mentioned in the Introduction, the mass of the $\Delta^{-}$ has
never been directly measured, we can either treat the relations as predictions
of its mass, or we can eliminate it from three of the relations using the
fourth; since all four of the relations result from chiral perturbation theory
alone, any linear combination of them is also a valid relation.  We choose to
eliminate the $\Delta^{-}$ using Eq.~\ref{first} (and its loop corrections),
because it is isospin-breaking and is the only one involving $\Delta$ masses
alone.  Because the other three relations depend only on measured quantities,
we obtain three nontrivial tests of HBEFT.

One problem that immediately crops up is that the $\Delta$ masses, as
presented in the Particle Data Group's (PDG) {\it Review of Particle Properties}
\cite{PDG}, rely on data fifteen years old, which generally have
substantial uncertainties relative to the isospin breaking of the multiplet.
The statistical averages of the accepted independent measurements in the PDG are
\begin{eqnarray} \label{pdg}
\Delta^{++} & = & 1230.86 \pm 0.13 \mbox{ MeV}, \nonumber \\
\Delta^{+}  & = & 1234.9  \pm 1.4  \mbox{ MeV}, \nonumber \\
\Delta^{0}  & = & 1233.42 \pm 0.16 \mbox{ MeV}.
\end{eqnarray}
We may ask about the reliability of these data.  A recent discussion of the
status of baryon isospin splitting measurements is found in a paper by
Cutkosky\cite{Cut}; in particular, the author points out that, in a quark model
fit, it is very difficult to accommodate the PDG values for the $\Delta$-mass
splittings.  One explanation is, of course, that the quark model is inadequate;
the other is that the measurements need to be refined.  The Virginia Polytechnic
Institute (VPI) group\cite{Arndt} currently recommends the value
\begin{equation} \label{vpi}
\Delta^0 - \Delta^{++} = 1.3 \pm 0.5 \mbox{ MeV}.
\end{equation}
The uncertainty depends on the measurement of scattering lengths and the
pion-nucleon coupling constant, and is expected to fall as their fit is refined.
Even so, this value with its current uncertainty is in disagreement with
Eq.~\ref{pdg}.

It would also be very helpful to bring down the large error in the PDG
$\Delta^+$ mass measurement.  There are a few pieces of information,
experimental and theoretical, in disagreement with the value in Eq.~\ref{pdg}.
This particular number is based on one measurement\cite{Mir}, and is in
discrepancy with three other independently measured PDG values, which have the
statistical average
\begin{equation} \label{dp}
\Delta^+ = 1231.5 \pm 0.3 \mbox{ MeV} .
\end{equation}
These were not used in the PDG fit because uncertainties of the individual
measurements were not estimated; the uncertainty given here is the statistical
variance, not experimental uncertainty.  However, we may use this information
as an alternative to the PDG value to demonstrate dependence of the results on
the measurement of $\Delta$ masses.  To support this choice, there are
additional predictions of a smaller $\Delta^+$ mass:  Morpurgo\cite{Morp}, using
his general parametrization method, predicts
\begin{eqnarray}
\Delta^+ & = & \Delta^{++} - (p - n) - (\Sigma^+ - 2\Sigma^0 + \Sigma^- ) \\
& = & 1230.45 \pm 0.27 \mbox{ MeV},
\end{eqnarray}
where we have used the PDG number for the $\Delta^{++}$ mass.  Furthermore,
pion-deuterium scattering data taken by Pedroni {\it et al.\/}\cite{Ped}
produce the mass combination measurement
\begin{equation}
\left( \Delta^- - \Delta^{++} \right) + \frac{1}{3} \left( \Delta^0 - \Delta^+
\right) = 4.6 \pm 0.2 \mbox{ MeV}.
\end{equation}
However, the error is statistical only and does not reflect a number of
theoretical corrections made in the processing of the data.  If we combine this
number with the relation Eq.~\ref{first}, the PDG value of $\Delta^{++}$, and
the VPI result, we find
\begin{equation}
\Delta^+ = 1230.78 \pm 0.52 \mbox{ MeV}.
\end{equation}
Because we are using the relation Eq.~\ref{first}, this is only a prediction,
rather than a true piece of data.  Nevertheless, these numbers all appear to be
roughly consistent and quite different from the PDG value.

In summary, to demonstrate the dependence of results on $\Delta$ masses, we
will exhibit our results twice:  once using only PDG numbers (data set A), and
once using the PDG number for $\Delta^{++}$ in Eq.~\ref{pdg}, the VPI result in
Eq.~\ref{vpi}, and the alternate value for the $\Delta^+$ mass given in
Eq.~\ref{dp} (data set B).

Another problem is the accuracy with which the axial-current couplings $\cal C$
and $\cal H$ are measured.  $\cal C$ is measured from strong
decuplet-to-octet baryon decays, and is reasonably well-known, but $\cal H$ can
occur only in decuplet-to-decuplet transitions, and thus is hard to determine.
The most current values available are\cite{Butler}
\begin{equation}
| {\cal C} | = 1.2 \pm 0.1, \hspace{2em} {\cal H} =  -2.2 \pm 0.6 .
\end{equation}
Fortunately, as we see in the Appendix, the leading terms dependent upon $\cal
H$ are actually $o(m_q^4 \ln m_q )$, and thus may be neglected in this model. 

A final point is that the expressions in general depend upon the decuplet and
octet baryon masses, the meson octet masses, the axial-current couplings, and
the quark mass parameter $r$.
As pointed out in the Appendix, this last factor arises from a consistent
treatment of $\pi^0$-$\eta$ and $\Sigma^0$-$\Lambda$ mixing; the question of its
exact size is open to debate, being related to the very interesting question of
whether the up-quark can be massless\cite{LL}.  All that can be said with
certainty is that $r = 0.03 \pm 0.02$, which accommodates all currently
suggested up-quark masses.  We find, however, that the terms containing $r$ are
numerically almost always small.

The results are presented as follows: for each set of $\Delta$ masses, we have a
prediction for the mass of the $\Delta^-$ (which we see in the Appendix to be
uncorrected to $o(m_q^{5/2})$), followed by the comparison of the
explicit breakings of the relations Eqs.~\ref{first}--\ref{last} (using
Eq.~\ref{first} to eliminate $\Delta^-$ dependence), and then
computed loop expressions for the corresponding breakings.
Experimental uncertainties of the decuplet masses are included.  For data set
A (all numbers in this Section in MeV),
\begin{equation}
\Delta^-  =  1226.42 \pm 4.23 ;
\end{equation}
\begin{eqnarray}
\left. \begin{array}{rcl}
\Delta_2 + \Delta_1 & = & -16.24 \pm 7.01 , \\
{\cal C}^2 \left[ (+1.26 - 1.4 r ) \pm 1.74 \right] & = & \; \: +1.88 \pm 2.53 ;
\end{array} \right\} & & \\
\left. \begin{array}{rcl}
\Delta_3 & = & \; \: +2.68 \pm 1.69 , \\
{\cal C}^2 \left[ (-0.01 + 20.5 r ) \pm 0.43 \right] & = & \; \: +0.87
\pm 0.87 ;
\end{array} \right\} & & \\
\left. \begin{array}{rcl}
\Delta_4 + \frac{1}{4} \Delta_1 & = & \; \: +5.47 \pm 1.75 , \\
{\cal C}^2 \left[ (-0.61 - 0.4 r ) \pm 0.25 \right] & = & \; \: -0.89 \pm 0.39 ;
\end{array} \right\} & &
\end{eqnarray}
and for data set B,
\begin{equation}
\Delta^- = 1232.84 \pm 1.81 ;
\end{equation}
\begin{eqnarray}
\left. \begin{array}{rcl}
\Delta_2 + \Delta_1 & = & -5.16 \pm 4.50 , \\
{\cal C}^2 \left[ (-0.70 + 0.5 r ) \pm 0.90 \right] & = & -0.98 \pm 1.30 ; \\
\end{array} \right\} & & \\
\left. \begin{array}{rcl}
\Delta_3 & = & +0.54 \pm 1.11 , \\
{\cal C}^2 \left[ (+0.19 - 18.9 r ) \pm 0.34  \right] & = & -0.54 \pm 0.74 ; \\
\end{array} \right\} & & \\
\left. \begin{array}{rcl}
\Delta_4 +\frac{1}{4} \Delta_1 & = & +5.91 \pm 1.69 , \\
{\cal C}^2 \left[ (+1.85 - 0.4 r ) \pm 0.59 \right] & = & +2.64 \pm 0.95 .
\end{array} \right\} & &
\end{eqnarray}

We see that, for both cases, the relation breakings are numerically of the same
order as the calculated loop corrections, which indicates reliability of the
loop expansion.  As to the particular results, data set B appears to give a
better fit, but to be certain we must estimate the size of higher-order
corrections in the theory.  Beyond what we have calculated, the next
contributions are two-loop effects of formal orders $m_q^{5/2}$ and
$m_q^3 \ln m_q$, and then the tree-level terms at $o(m_q^3)$.
We expect the two-loop corrections to be numerically small compared to the
one-loop contributions, so we are led to consider the size of third-order
contributions to Eqs.~\ref{first}--\ref{last}.  There are only two nontrivial
such terms:
\begin{equation}
\frac{1}{\Lambda_\chi} \overline{T}^{\mu}_{ijk} M^{i}{}_{l} M^{j}{}_{m}
M^{k}{}_{n} T_{\mu}^{lmn} , \; \;
\frac{\alpha}{4 \pi} \overline{T}^{\mu}_{ijk} M^{i}{}_{l} Q^{j}{}_{m}
Q^{k}{}_{n} T_{\mu}^{lmn} .
\end{equation}
We may now estimate the generic third-order contributions to the relations
using the isospin transformation properties of $\Delta_{1,2,3,4}$, because the
$\Delta I = 1$ portion of $M_q$ is proportional to the small parameter $r$.
With a unit-size chiral coefficient, $r \sim 0.03$, $\Lambda_\chi \sim$ 1
GeV, $m_s \sim$ 200 MeV, we have the following estimates for $o(M^3)$ and
$o(MQ^2)$ contributions, respectively:
\begin{eqnarray}
\Delta_1 : & 2 \cdot 10^{-4}, & 3 \cdot 10^{-3} , \nonumber \\
\Delta_2 : & 7 \cdot 10^{-3}, & 0.1 , \nonumber \\
\Delta_3 : & 0.2            , & 0.1 , \nonumber \\
\Delta_4 : & 8              , & 0.1 .
\end{eqnarray}
We see that these contributions are the right numerical order to explain the
differences between the one-loop contributions and the experimental breakings,
with the exception of the quantity $(\Delta_2 + \Delta_1)$ for data set A; this
leads us to favor data set B.  Another distinction between the two sets is the
prediction for the $\Delta^-$ mass; once it is measured, it will probe the
validity of the prediction of this model that $\Delta_1 = 0$ to $o(m_q^{5/2})$.

We can also fit to the ten chiral coefficients listed in Eq.~\ref{mass}; with
nine known decuplet masses, one parameter remains, which we chose to be
$c^{64}_3$.  Because we have seen that the relation Eq.~\ref{first} remains
unbroken by the loop contributions we have considered, and the third-order
tree-level contributions are estimated to be tiny, we expect $c^{64}_3$ to be
quite small ($\ll$ 0.1 MeV).  With data set B, we obtain the following fit:
\begin{equation}
\begin{array}{ll}
\begin{array}{lcrl}
c^1_0    & = & 1382.03 \pm 0.24 & -            2 c^{64}_3 , \\
c^8_0    & = & -148.43 \pm 0.21 & -            2 c^{64}_3 , \\
c^8_1    & = &   -1.24 \pm 0.38 & +            4 c^{64}_3 , \\
c^{27}_0 & = &   -0.64 \pm 0.04 & - \frac{2}{7}  c^{64}_3 , \\ 
c^{27}_1 & = &   +0.28 \pm 0.09 & + \frac{6}{7}  c^{64}_3 , \\
\end{array} &
\begin{array}{lcrl}
c^{27}_2 & = &   +0.06 \pm 0.15 & - \frac{10}{7} c^{64}_3 , \\
c^{64}_0 & = &   +0.17 \pm 0.05 & - \frac{1}{7}  c^{64}_3 , \\
c^{64}_1 & = &   +0.01 \pm 0.03 & + \frac{1}{7}  c^{64}_3 , \\
c^{64}_2 & = &   -0.18 \pm 0.16 & - \frac{5}{7}  c^{64}_3 .
\end{array}
\end{array}
\end{equation}
The interesting feature here is that the uncertainties in the decuplet mass
differences are so large that it becomes increasingly difficult to determine
reliably the coefficients of the larger $SU(3)$ representations.  Precise
measurements of baryon decuplet mass differences would allow us to pin down
these coefficients and thus constrain those particular operators with the same
transformation properties.

\section{Conclusions}
We have seen that the heavy baryon chiral perturbation theory predicts four
nontrivial relations between the decuplet masses which hold to second order in
flavor breaking.  That exactly this number of relations must occur at this order
was explained in terms of parameter counting in the Lagrangian and the
group-theoretical properties of possible operators.  We have observed that any
relations holding to second order have calculable, renormalization-point
independent corrections, and we have computed and tabulated these corrections.
Finally, we have seen that the experimental values of the breaking of these
relations can be accounted for by the one-loop chiral corrections, and we have
argued that higher-order effects are not so large as to spoil this fit.  In
particular, the relation Eq.~\ref{first} is seen to be exact to the limits of
this model, and thus one can predict the mass of the $\Delta^-$ using the other
$\Delta$ masses.

When, in the future, the experimental values for the decuplet mass splittings
and their current couplings are refined, the expressions derived in this paper
will provide four sensitive checks on heavy baryon chiral perturbation theory.
For the moment, however, this simple theory appears to be quite satisfactory in
explaining the current experimental limits.

\section*{Acknowledgements}
I would like to thank Markus Luty for his suggestions and comments on the
manuscript, Charles Wohl and Richard Arndt for their expert information on the
$\Delta$ resonance masses, and Mahiko Suzuki for valuable discussions on
group-theoretical aspects of this work.  This work was supported in part by the
Director, Office of Energy Research, Office of High Energy and Nuclear Physics, 
Division of High Energy Physics of the U.S. Department of Energy under Contract
DE-AC03-76SF00098 and in part by the National Science Foundation under grant
PHY-90-21139.

\section*{Appendix}
Here we present important details in the calculation of loop diagrams to enable
the reader to understand the nature of the computations.  We also present the
full expressions for the one-loop corrections to the decuplet mass relations
Eqs.~\ref{first}--\ref{last}.

As discussed in Section 6, we will need to compute only those diagrams with two
trilinear vertices, since keyhole diagrams respect the relations.  The mass
contribution, as computed using the Feynman rules of HBEFT as outlined in
Ref.~\cite{Jenk3}, is obtained from the generic loop expression
\begin{eqnarray}
\lefteqn{\Sigma ( p_i \cdot v ) =} \nonumber \\
& & \hspace{-1em} \mbox{} +\frac{i}{f^2} \mbox{({\it Clebsch})}_{i,j,\alpha}^2
\overline{T}^{\mu} \left\{ \int \frac{d^4 k}{\left( 2 \pi \right)^4}
\frac{1}{\left( k + p_i \right) \cdot v  - \Delta m_j } \frac{1}{k^2 -
m_{\alpha}^2} k_{\mu} k^{\nu} \right\} T_{\nu},
\end{eqnarray} 
where the index $i$ refers to a particular external decuplet state, $j$ refers
to the intermediate baryon, and $\alpha$ refers to the loop meson, with an
implicit sum over the latter two indices.  The external momentum $p_i$ is
residual, the common octet mass $m_8$ having been removed through HBEFT.  The
group-theoretical couplings {\it Clebsch} are readily computed by constructing
matrix representations of the generators for the octet meson and baryon and
decuplet baryon states (as in Eqs.~\ref{m1}--\ref{m3}) and taking appropriate
traces.  There are also suppressed factors from contraction of spin and
projection operators and the couplings ${\cal H}^2$, ${\cal C}^2$
for decuplet- and octet-mediated diagrams, respectively.

The term $\Delta m_j$ is designed to implement our program of including full
physical baryon masses instead of chiral coefficients for internal lines.  If
we worked only with chiral coefficients, this term would read $\Delta m, 0$
for decuplet(octet) intermediates.  Instead, let $\delta m$ indicate the amount
by which a baryon exceeds its multiplet's common mass.  Then, in the rest
frame, $p_i^{\mu} = ( \delta m_i + \Delta m, \bf{0} )$ and
$v^{\mu} = (1, \bf{0})$, so that $\, p_i \cdot v = \delta m_i + \Delta m$ in all
frames.  For decuplet(octet) intermediates, $\Delta m_j = ( \delta m_j +
\Delta m ),
\delta m_j$, respectively.  As a result, we find that the integral depends
on baryon masses through their differences only, for then $p_i \cdot v -
\Delta m_j = m_i - m_j$.  One further point is that
the mass contribution will issue from $-\Sigma(p_i \cdot v)$ with the external
momentum as chosen above, where the extra sign comes from the fact that the
spin-3/2 kinetic term has the opposite sign to that for the usual spin-1/2 case.

There is a small amount of sleight of hand here, for had we used the decuplet
masses for intermediate lines from the outset, we would not have obtained the
cancellation of the diagrams as described in Section 6 with ${\cal O} = \bf{1}$
(the $o(m_q^{3/2})$ corrections).  Instead, inclusion of the full decuplet
masses would mix the ${\cal O} = \bf{1}$ terms with all the others.  But this is
exactly what we do with the octet-mediated diagrams.  The difference is the
presence of the octet-decuplet splitting parameter $\Delta m$ in the latter.
Although experimentally {\it intra}multiplet splittings may be of the same
magnitude as {\it inter}multiplet splittings, they are formally two different
phenomena in the chiral Lagrangian, where the octet and decuplet are taken as
independent.

We then find the following expressions for $\delta m_i$:
\begin{equation} \label{h2}
\delta m_i = \frac{5}{6} \mbox{({\it Clebsch})}_{i,j,\alpha}^2
\frac{{\cal H}^2}{16 \pi^2 f^2} m_{\alpha}^2 \ln \left(
\frac{m_{\alpha}^2}{\mu^2} \right) \left( m_j - m_i \right) ,
\end{equation}
for decuplet-mediated diagrams, and
\begin{equation}
\delta m_i = \frac{1}{3} \mbox{({\it Clebsch})}_{i,j,\alpha}^2
\frac{{\cal C}^2}{16 \pi f^2} m_{\alpha}^3 \ H ( \xi^{\alpha}_{ij} ),
\end{equation}
for octet-mediated diagrams, where
\begin{equation}
\xi^{\alpha}_{ij} \equiv \frac{m_i - m_j}{m_{\alpha}},
\end{equation}
\begin{eqnarray}
F(\xi) & \equiv & \int_0^{\infty} dx \, \left( x^2 + 2 \xi x + 1 \right)^{-1}
\nonumber \\ & = & \left\{
\begin{array}{lc}
\frac{1}{\sqrt{\xi^2 - 1}} (\mbox{sgn} \, \xi ) \cosh^{-1} \left| \xi \right| ,
& \left| \xi \right| > 1 \vspace{1ex} \\
\frac{1}{\sqrt{1 - \xi^2}} \cos^{-1} \xi , & \left| \xi \right| < 1
\end{array} \right. ,
\end{eqnarray}
and
\begin{equation}
H ( \xi ) \equiv \frac{3}{2 \pi} \left[ \xi \left( \frac{2}{3} \xi^2 - 1
\right) \ln \frac{m_{\alpha}^2}{\mu^2} - \frac{4}{3} \left( \xi^2 -1 \right)^2
F( - \xi ) \right] .
\end{equation}

The corrections in Eq.~\ref{h2}, and indeed any corrections linear in the
decuplet masses, have a remarkable property:  {\em a priori} they are of order
$m_q^2 \ln m_q$, but in fact they are of order $m_q^4 \ln m_q$, and are thus
formally smaller than the third-order tree-level terms we are neglecting.  The
reason is that the relations Eq.~\ref{first}--\ref{last} are broken only by the
${\bf 64}$ representation of $SU(3)$, as discussed in Section 5.  Therefore, for
each octet meson $\alpha$, we have a linear combination of decuplet masses which
must be the coefficient of a ${\bf 64}$ operator constructed of the decuplet
fields.  But the only such combinations are exactly those in the combinations
$\Delta_{1,2,3,4}$, which are $o(m_q^3)$.  Thus we may neglect the corrections
in Eq.~\ref{h2} in this model.

Now we are ready to tabulate the breakings of the relations
Eqs.~\ref{first}--\ref{last}, which have been labeled $\Delta_{1,2,3,4}$.  As it
stands, however, the expressions will be quite cumbersome unless we compactify
our notation.  We denote
\begin{equation}
\frac{m_{\alpha}^3}{16 \pi f^2} \, H( \xi^{\alpha}_{ij} ) \rightarrow
\tilde{\alpha} (i,j) ,
\end{equation}
where the indices are replaced with the particles they represent.  We then have
\begin{eqnarray}
\lefteqn{\Delta_1 =} \nonumber \\ & & \hspace{-0.5em}
\frac{1}{3} {\cal C}^2 \left\{ \, + \tilde{\pi}^+ \left( ( \Delta^{++}, p) - 
( \Delta^+ , n) + ( \Delta^0 , p) - ( \Delta^- , n) \right)
\right. \nonumber \\ & & \hspace{2em}
-2 \tilde{\pi}^0 \left( ( \Delta^+ , p) - (
\Delta^0 , n) \right)
\nonumber \\ & & \hspace{2em}
+ \tilde{K}^+ \left( ( \Delta^{++}, \Sigma^+ ) - 2 ( \Delta^+ , \Sigma^0 ) 
+ ( \Delta^0 , \Sigma^- ) \right)
\nonumber \\ & & \hspace{2em} \left. \!
- \tilde{K}^0 \left( ( \Delta^{+}, \Sigma^+ ) - 2 ( \Delta^0 , \Sigma^0 ) 
+ ( \Delta^- , \Sigma^- ) \right) \right\} , \nonumber \\ & &
\end{eqnarray}
\begin{eqnarray}
\lefteqn{\Delta_2 =} \nonumber \\ & & \hspace{-1em}
\frac{1}{3} {\cal C}^2 \left\{ + \frac{1}{3} \tilde{\pi}^+ \left[ \,
3( \Delta^{++} , p) - ( \Delta^+ , n) - ( \Delta^0 , p) + 3( \Delta^- , n)
\right. \right. \nonumber \\ & & \hspace{5em}
- \left( 1 - \frac{3}{2} r \right) ( \Sigma^{*+}, \Sigma^0 )
-3 \left( 1 - \frac{1}{2} r \right) ( \Sigma^{*+}, \Lambda )
\nonumber \\ & & \hspace{5em}
+ 2 \left( ( \Sigma^{*0}, \Sigma^+ ) + ( \Sigma^{*0}, \Sigma^- ) \right)
\nonumber \\ & & \hspace{5em} \left.
- \left( 1 - \frac{3}{2} r \right) ( \Sigma^{*-}, \Sigma^0 )
-3 \left( 1 + \frac{1}{2} r \right) ( \Sigma^{*-}, \Lambda )
\right] \nonumber \\ & & \hspace{1.5em}
- \frac{1}{3} \tilde{\pi}^0 \left[ \, 2 \left(
( \Delta^+ , p) + ( \Delta^0 , n) \right) -6 ( \Sigma^{*0}, \Lambda ) \right.
\nonumber \\ & & \hspace{5em} \left.
+ \left( 1 + \frac{3}{2} r \right) ( \Sigma^{*+},
\Sigma^{+} ) + \left( 1 - \frac{3}{2} r \right)
( \Sigma^{*-}, \Sigma^{-} ) \right]
\nonumber \\ & & \hspace{1.5em}
- \hspace{0.5em} \tilde{\eta} \hspace{1em} \left[ - 2 ( \Sigma^{*0}, \Sigma^0 )
+ \left( 1 - \frac{1}{2} r \right) (\Sigma^{*+}, \Sigma^+ ) +
\left( 1 + \frac{1}{2} r \right) ( \Sigma^{*-}, \Sigma^- )
\right]
\nonumber \\ & & \hspace{1.5em}
+ \frac{1}{3} \tilde{K}^+ \! \left[ \, 3 ( \Delta^{++}, \Sigma^+ )
- 2 ( \Delta^+ , \Sigma^0 ) - ( \Delta^0 , \Sigma^- ) \right.
\nonumber \\ & & \hspace{5em} \left.
-2 \left( (\Sigma^{*+}, \Xi^0 ) - (\Sigma^{*0}, p) - (\Sigma^{*0}, \Xi^- ) +
(\Sigma^{*-}, n ) \right) \right]
\nonumber \\ & & \hspace{1.5em}
- \frac{1}{3} \tilde{K}^0 \left[ (\Delta^+ , \Sigma^+ ) + 2 ( \Delta^0 ,
\Sigma^0 ) - 3 (\Delta^- , \Sigma^- ) \right.
\nonumber \\ & & \hspace{5em} \left.
+2 \left( (\Sigma^{*+}, p) - (\Sigma^{*0}, n) - (\Sigma^{*0}, \Xi^0 ) +
(\Sigma^{*-}, \Xi^- ) \right) \right] , \nonumber \\ & &
\end{eqnarray}
\begin{eqnarray}
\lefteqn{\Delta_3 =} \nonumber \\ & & \hspace{-1em}
\frac{1}{3} {\cal C}^2 \left\{ +\frac{1}{6} \tilde{\pi}^+ \hspace{1em}
\left[ \, 2 \left(
(\Delta^+ , n) - (\Delta^0 , p) + (\Xi^{*0}, \Xi^- ) - (\Xi^{*-}, \Xi^0 )
\right) \right. \right.
\nonumber \\ & & \hspace{5.5em}
- \left( 1 + \frac{3}{2} r \right) (\Sigma^{*+}, \Sigma^0 ) -
3 \left( 1 - \frac{1}{2} r \right) (\Sigma^{*+}, \Lambda )
\nonumber \\ & & \hspace{5.5em} \left.
+ \left( 1 - \frac{3}{2} r \right) (\Sigma^{*-}, \Sigma^0 ) +
3 \left( 1 + \frac{1}{2} r \right) (\Sigma^{*-}, \Lambda )
\right] \nonumber \\ & & \hspace{1.5em}
+ \frac{1}{6} \tilde{\pi}^0 \hspace{1.2em} \left[ \, 4
\left( (\Delta^+ , p) - (\Delta^0 , n) \right) \right.
\nonumber \\ & & \hspace{5.5em}
- \left( 1 + \frac{3}{2} r \right) \left( (\Sigma^{*+},
\Sigma^+ ) - (\Xi^{*0}, \Xi^0 ) \right)
\nonumber \\ & & \hspace{5.5em} \left.
+ \left( 1 - \frac{3}{2} r \right) \left( (\Sigma^{*-},
\Sigma^- ) - (\Xi^{*-}, \Xi^- ) \right) \right]
\nonumber \\ & & \hspace{2em}
+ \frac{1}{2} \tilde{\eta} \hspace{1em} \left[ 
- \left( 1 - \frac{1}{2} r \right) \left( (\Sigma^{*+},
\Sigma^+ ) - ( \Xi^{*0}, \Xi^0 ) \right) \right.
\nonumber \\ & & \hspace{5.5em} \left.
+ \left( 1 + \frac{1}{2} r \right) \left( (\Sigma^{*-},
\Sigma^- ) - (\Xi^{*-}, \Xi^- ) \right) \right]
\nonumber \\ & & \hspace{2em}
+ \frac{1}{6} \tilde{K}^+ \left[ \, 4
(\Delta^+ , \Sigma^0) + \right.
\nonumber \\ & & \hspace{5.5em} 
-2 \left( (\Delta^0 , \Sigma^- ) + (\Sigma^{*+}, \Xi^0 ) - (\Sigma^{*-}, n) -
(\Xi^{*0}, \Sigma^+ ) \right)
\nonumber \\ & & \hspace{5.5em} \left.
- \left( 1 - \frac{3}{2} r \right) (\Xi^{*-}, \Sigma^0 ) - 3
\left( 1 + \frac{1}{2} r \right) (\Xi^{*-}, \Lambda ) \right]
\nonumber \\ & & \hspace{2em}
- \frac{1}{6} \tilde{K}^0 \left[ \, 4 (\Delta^0, \Sigma^0 )
-2 \left( (\Delta^+ , \Sigma^+ ) - (\Sigma^{*+}, p) + (\Sigma^{*-}, \Xi^- ) -
(\Xi^{*-}, \Sigma^- ) \right) \right.
\nonumber \\ & & \hspace{5.5em} \left.
- \left( 1 + \frac{3}{2} r \right) (\Xi^{*0}, \Sigma^0 ) -
3\left( 1 - \frac{1}{2} r \right) (\Xi^{*0}, \Lambda ) \right]
, \nonumber \\ & &
\end{eqnarray}
\begin{eqnarray}
\lefteqn{\Delta_4 =} \nonumber \\ & & \hspace{-1em}
\frac{1}{3} {\cal C}^2 \left\{ +\frac{1}{12} \tilde{\pi}^+ \left[ \,
3 (\Delta^{++}, p) + (\Delta^+ , n) + (\Delta^0 , p) + 3 (\Delta^- , n) \right.
\right.
\nonumber \\ & & \hspace{5em}
- 2 \left( 1 + \frac{3}{2} r \right) (\Sigma^{*+}, \Sigma^0 )
- 6 \left( 1 - \frac{1}{2} r \right) (\Sigma^{*+}, \Lambda )
\nonumber \\ & & \hspace{5em}
- 2 \left( ( \Sigma^{*0}, \Sigma^+ ) + (\Sigma^{*0}, \Sigma^- ) \right)
\nonumber \\ & & \hspace{5em}
- 2 \left( 1 - \frac{3}{2} r \right) (\Sigma^{*-}, \Sigma^0 )
- 6 \left( 1 + \frac{1}{2} r \right) (\Sigma^{*-}, \Lambda )
\nonumber \\ & & \hspace{5em} \left.
+ 6 \left( (\Xi^{*0}, \Xi^- ) + (\Xi^{*-}, \Xi^0 ) \right) \right]
\nonumber \\ & & \hspace{1.5em}
+ \frac{1}{12} \tilde{\pi}^0 \left[ \, 2 \left(
(\Delta^+ , p) + (\Delta^0 , n) \right) - 6 (\Sigma^{*0}, \Lambda) \right.
\nonumber \\ & & \hspace{5em}
- \left( 1 + \frac{3}{2} r \right) \left( + 2 (\Sigma^{*+},
\Sigma^+ ) - 3 (\Xi^{*0}, \Xi^0 ) \right)
\nonumber \\ & & \hspace{5em} \left.
- \left( 1 - \frac{3}{2} r \right) \left( + 2 (\Sigma^{*-},
\Sigma^- ) - 3 (\Xi^{*-}, \Xi^- ) \right) \right]
\nonumber \\ & & \hspace{1.5em}
+ \frac{1}{4} \hspace{0.5em} \tilde{\eta} \hspace{0.5em} \left[ 
- 2 ( \Sigma^{*0}, \Sigma^0 ) \right.
\nonumber \\  & & \hspace{5em}
- \left( 1 - \frac{1}{2} r \right) \left( + 2 (\Sigma^{*+},
\Sigma^+ ) - 3 (\Xi^{*0}, \Xi^0 ) \right)
\nonumber \\ & & \hspace{5em} \left.
- \left( 1 + \frac{1}{2} r \right) \left( + 2 (\Sigma^{*-},
\Sigma^- ) - 3 (\Xi^{*-}, \Xi^- ) \right) \right]
\nonumber \\ & & \hspace{1.5em}
+ \frac{1}{12} \tilde{K}^+ \! \left[ \, 3 (\Delta^{++} \! , \Sigma^+ ) + 2 
(\Delta^+ , \Sigma^0 ) + (\Delta^0 , \Sigma^- ) \right.
\nonumber \\ & & \hspace{5em}
- 2 \left( 2 (\Sigma^{*+}, \Xi^0 ) + (\Sigma^{*0}, p ) + (\Sigma^{*0}, \Xi^- ) +
2 ( \Sigma^{*-}, n ) \right)
\nonumber \\ & & \hspace{5em}
+ 3 \left( \left( 1 - \frac{3}{2} r
\right) (\Xi^{*-}, \Sigma^0 ) + 9 \left( 1 + \frac{1}{2} r
\right) (\Xi^{*-}, \Lambda ) \right)
\nonumber \\ & & \hspace{5em} \left.
+ 6 \left( (\Xi^{*0}, \Sigma^+ ) - 2 (\Omega^{-}, \Xi^0 ) \right) \right]
\nonumber \\ & & \hspace{1.5em}
+ \frac{1}{12} \tilde{K}^0 \left[ (\Delta^+ , \Sigma^+ ) + 2 
(\Delta^0 , \Sigma^0 ) + 3 ( \Delta^- , \Sigma^- ) \right.
\nonumber \\ & & \hspace{5em}
- 2 \left( 2 (\Sigma^{*+}, p ) + (\Sigma^{*0}, n ) + (\Sigma^{*0}, \Xi^0 ) + 2
(\Sigma^{*-}, \Xi^- ) \right)
\nonumber \\ & & \hspace{5em}
+ 3 \left( \left( 1 + \frac{3}{2} r \right) (\Xi^{*0},
\Sigma^0 ) + 9 \left( 1 - \frac{1}{2} r \right) (\Xi^{*0},
\Lambda ) \right)
\nonumber \\ & & \hspace{5em} \left. \left.
+ 6 \left( (\Xi^{*-}, \Sigma^- )  - 2 (\Omega^- , \Xi^- ) \right) \right]
\right\} . \nonumber \\ & &
\end{eqnarray}
The factors of $r$ arise from $\pi^0$-$\eta$ and $\Sigma^0$-$\Lambda$ mixing,
which we incorporate into the calculation by rotating the $SU(3)$ generators to
$o(r)$; this eliminates the mass-mixing terms to order ($o(r^2)$), consistent
with the model.  However, within decuplet loop corrections, the $o(r^2)$ terms
will be numerically insignificant, so they have been suppressed in the above
expressions.

In Section 6 we indicated that the octet loop contributions
would vanish if we took the value of {\it all} decuplet-octet splittings to be
$\Delta m$; in that case, $\xi^{\alpha}_{ij} = \Delta m / m_{\alpha}$ depends
only on $\alpha$, the meson index.  Thus we expect the coefficient of each
$\tilde{\pi}^{\alpha}$ to vanish in the loop corrections if we replace each
$\tilde{\alpha} (i,j)$ with the same factor $f(\alpha)$, as can be trivially
verified. 

Upon expanding all meson and baryon masses in terms of quark masses and charges,
we find that the ${\cal C}^2$ term in $\Delta_1$ is also formally too small
($o(m_q^3 \ln m_q)$) to keep in the current calculation, because there are
two-loop effects at $o(m_q^{5/2})$ which have been neglected.  This is a result
of the fact that $\Delta_1$ comes from $\Delta I = 3$ terms, which require three
powers of quark masses, of which at least two powers in the loop contributions
must come from the decuplet masses.  Thus we find in this model that
$\Delta_1 = 0$, so that the tree-level relation is uncorrected.

\newpage
\section*{Figure captions}
FIG. 1a.  ``Keyhole'' (quartic vertex) diagram contributing to baryon masses.

\noindent
FIG. 1b.  Trilinear vertex diagram contributing to baryon masses.  The internal
baryon  line may be either octet or decuplet.
\end{document}